Familiar biological, chemical and physical events credibly evolve the Standard Genetic Code

Michael Yarus

Department of Molecular, Cellular and Developmental Biology
University of Colorado  Boulder
Boulder, Colorado 80309-0347
Phone: +1 303 817-6018
Email: yarus@colorado.edu

**Abstract**

The genetic code is profoundly shaped by an origin in ancient RNA-mediated interactions, needing an extended development to reach the Standard Genetic Code (SGC). That development can serially use RNA specificities, a ribonucleopeptide transition (RNPT), finally code escape and diaspora. An index of evolutionary plausibility based on least selection takes simultaneous account of speed and accuracy of evolution, identifying favored evolutions. Combining RNA world specificities allowed convergence of early coding to SGC assignments. Secondly, this was sufficient to launch a post-RNA-world RNPT. The RNPT allowed biosynthesis of complex amino acids, depending heavily on late code fusions between coexisting independent codes. Thirdly, escape from fluctuating, but highly-evolved codes of the RNPT applied a near-ideal selection for fastest-evolving and most accurate/useful genetic codes. Concurrently, a code and its microbial carrier suited to a free-living existence necessarily evolved. The established unity of life on Earth likely traces to SGC ascendancy during escape from the RNPT, and code diaspora.

Introduction

**The route to the code.** This manuscript discusses origins of the Standard Genetic Code, whose closely related form in every known Earth organism (1) strongly implies a common Last Universal Common Ancestor for existing Earth biota (LUCA, 2). Because unaided RNA-based coding appears sufficiently complete to initiate evolution to the later, more competent SGC (3), it seems likely that an earlier era of directly-RNA-encoded peptides (4, 5) was followed by a ribonucleopeptide transition (RNPT) featuring enhanced coding. The enhancements featured more complex amino acids biosynthesized using peptide-aided catalysis, and these catalysts were also encoded on a more complex translation apparatus (6), composed of both active RNAs and peptides (7).

**Calculation of evolutionary routes.** Appearance of the SGC during the RNPT is analyzed, using a quantitative kinetic model (8) see in which large numbers of codes are followed though evolutionary time in numerous independent code environments (Methods). Environments each allow code evolution events (for example, triplet assignment, decay, codon capture, code fusion….) at specified rates. Code events are assigned probabilities during a brief interval, called a passage, small enough that only one event occurs during a passage. This is mathematically equivalent to assigning differing chemical rate constants to these varied evolutionary events (8); thus explicit time courses for multi-event code evolution pathways are computable.

Thus probable kinetics of SGC appearance during the RNA world or RNPT can be calculated, and more rapid and accurate evolution can be explicitly defined. Specifically, the path of least selection (9) is sought; that is, the evolutionary route that most rapidly brings a population of codes into close proximity to the SGC.

**Shape of the assignments.** Code evolution is unexpectedly dependent on the shape and size of elements that are gathered into a finite coding table (10). In particular, if coding grows by assigning units that are larger or irregular in size, code evolution slows. As a specific example, if coding is by triplets that immediately wobble, acquisition of complete coding is significantly slowed (11) where "complete" has the meaning "having all functions encoded". Accordingly, early wobble is notably sub-optimal (11). Initial assignments were likely individual triplets, with wobble delayed until most assignments had been made.

**The immediate issue.** Coding most plausibly begins with the most accessible system that can mediate specific relations between RNA or RNA-like sequences and amino acids. Said another way, a complex coding apparatus cannot suddenly appear, but must progressively arise from simpler precursors that are also continuously functional. This introduction makes clear the importance of specific molecular interactions between oligoribonucleotides and amino acids (11); their multiply-demonstrated existence (12, 13) makes plausible a primordial translation system composed of only two molecules; activated amino acids and RNA. This early direct RNA ordering of short peptides can then evolve to add more complex amino acids and a more sophisticated proto-ribosome: in the example above, one that enables wobble coding.

Simple initial coding with RNA and early activated amino acids alone requires a transition to a more capable system. A crucial finding is that the breadth of amino acid affinities implied by existing measurements of RNA-amino acid interaction (3) is sufficient to specify synthesis of more capable peptide catalysts. That this is possible has been demonstrated: existing highly-evolved protein enzymes can be reduced to a much smaller variety of amino acid constituents (14, 15). These are few enough to have been specifically ordered/encoded on RNA sequences (4, 5).

The SGC would then arise from two intermediate events: firstly, early simpler peptides complemented RNA to synthesize more complex amino acids. These ribonucleopeptides might themselves be sufficient to support anabolic pathways, but simple early RNA catalysts, such as cofactors (16), might also join to elaborate biosynthesis for the complex amino acids (17). NAD alone, for example, would have enabled synthesis of an almost modern variety of amino acids from primordial metabolic precursors (18). Notably, ribozymic biosynthesis of 5' NAD RNA has been demonstrated (19), and NAD readily initiates transcription and so occurs at the 5' of transcribed RNAs (20), making cofactor-ribozymes practical. Secondly, simple ancient peptides would join existing oligoribonucleotides (21) to make a more precise, more capable ribosome (7). These events together are the RNPT. Approximating this transition is a principal object of present calculations.

**Approximating the RNPT.** While RNA catalysis (22, 23) has proven to be far more versatile than initially expected (24), RNA is comparatively impoverished with regard to catalytic groups and thus commonly slower than protein catalysis. Thus, a salient quality of the RNPT is that it bridges two types of catalysts, beginning with RNA alone, but ending with RNA, RNA plus peptides, and independent peptide catalysts for biological reactions. This is significant, because RNA catalysis is experimentally estimated to be $10^3$-fold slower than protein catalysis of the same reaction (25). More rapid catalysts ultimately transform the biosphere, because cells with more rapid biosynthesis can divide more rapidly. More rapidly dividing cells quickly colonize the biosphere, and become unique progenitors of subsequent biota. The consequences of this accelerated catalysis during the RNPT are featured below.

**Modeling alteration of cell division by catalytic interconversion.** In Methods, the following is more explicitly described; cell division presumably requires synthesis of catalysts and structural guides for partition of cell contents. Their synthesis should be accelerated by peptide-mediated chemistry. But even after (bio)chemical provisions for division are accelerated, division is still rate-limited by slower macroscopic processes during cell separation (Fig. 1).

The model uses data from a nucleotide diphosphate kinase enzyme, capable of minimal catalysis when composed of 13 different amino acids (15). Fig. 1 shows the derived effect on cell division (Methods). Division shifts from its early RNA rate-limit, accelerates due to intrinsically $10^3$-fold faster peptide catalysis, but ends up at a second, now-physical, cell separation rate limit (reached when 20 functions are encoded).

**Relation to coding.** Code evolution is approximated in this way: when using the RNA catalysis limit for cell division, we are in the RNA world (26). After peptide catalysis becomes significant ($\geq$ 15 amino acids are encoded), we have entered the RNPT. There are small peptide effects at 15 encoded functions, then substantially faster cell divisions as near-complete (20 functions; Fig. 1) coding is approached. Discussion of encoded translation initiation and termination is deferred, given phylogenetic evidence that their mechanisms were settled later, after separation of biological domains (27).

**Evolution to 15 coded functions (RNA encoding) and 20 functions (RNPT).** In this work, evolutionary paths are compared on 2-dimensional graphs. This presentation (28) plots evolutionary outcomes (y-axis) versus a structured x-axis list of all possible mechanisms; effects traceable to specific mechanisms appear as periodicities along the x-axis. To illustrate this method, Fig. 2A presents a plot of 4-dimensional information about 16 kinds of RNA world evolution.

**The RNA era has $\leq$ 15 encoded functions**. Fig. 2A compares effects of constant division rates at the low RNA-era rate (**Pdiv**) with division that responds to the increasing activity of peptides (**DivProb**; Methods; Fig. 1; see **Approximating the RNPT,** above). Fig. 2A also distinguishes the effects of initiation with a single triplet evident in experimental RNA-amino acid binding sites (**1SGC**) from initiation with triplets at the frequency implied for entire coding tables from amino acid binding selections (**SGC_r3**; 3). SGC_r3 initiations therefore simulate an early, rapidly-formed stereochemical code.

The figure header also indicates tests of continual new code initiation (**tab**) alongside the initial evolving code, or alternatively, coding evolving in an environment with only one initial code (**notab**). Finally, Fig. 2A also evaluates the effects of code fusions (**fus**) alongside coding environments in which codes do not fuse (**nofus**). These distinctions were chosen because they previously produced significant effects: for example, multiple codes and code fusion purify coding, producing convergence to a consensus of prior code assignments (29). These mechanisms are evaluated in all possible combinations (Methods), so Fig. 2 and its sequels arguably offer a broad survey of paths to 15 function (RNA world) codes and 20 function (RNPT; Fig. 3) codes (below).

**RNA era: rate findings.** RNA-era times to 15-function codes (Fig. 2A) reveal that these pathway changes infrequently alter evolutionary speed. For example, the difference between code division unresponsive to code completeness (Pdiv) and responsive division (DivProb) is slight, and the effects of independently originating codes (tab) and none (notab) plus code fusions (fus) or none (no fus) are both also slight. Moreover, effects of parallel codes and fusion are similar, whether (DivProb) or not (Pdiv) code division is responsive to coding.

Substantial differences occur only in one case: when comparing codes originating with a single encoded triplet (1SGC) and then making successive single SGC-like assignments, or instead, acquiring many assignments early, effectively at once (SGC_r3), then making single assignments.

These differences, between ≈200 passages (1SGC) and ≈50 passages (SGC_r3) to evolve 15 functions, are attributable to the fact that SGC_r3 encodes 14 functions from the start (3). Thus early stereochemical code initiation by SGC_r3 provides a radically shorter evolution to 15 encoded functions. But note: responsive code division, continuous initiation of codes and code fusions only slightly alter 1SGC initiation's course (Fig. 2A), even though 1SGC initiated paths must evolve 14 new encodings during a 4-fold longer time (Fig. 2A). Thus, an RNA-world's duration is relatively insensitive to environmental path details (Fig. 2A).

**RNA era: accuracy findings.** The second requirement for least selection (9) is that evolution be accurate; requiring least additional change. The frequency of evolved codes identical with the SGC (mis0), or differing by only one assignment (mis1) are responsive indicators of coding accuracy (28). Fig. 2B plots accuracy as mis0 and mis1 for the same 16 paths as Fig. 2A.

Fig. 2B is, approximately an inverse version of Fig. 2A; fast paths also tend to be accurate ones (28). We return to speed and accuracy's relation below. But in Fig. 2B, longer paths (1SGC; Fig. 1A) also tend to be more exposed to, and aided by independent parallel codes (tab, notab) and code fusions (fus, nofus). The accuracy of these RNA-era codes is slightly aided by fusion and particularly, by coexisting codes (29). But refinement by interaction with environmental codes is suppressed by more inclusive code initiation (SGC_3, Fig. 2B).

**Summary of RNA world codes: an index of plausibility.** Given that the most probable evolutionary path reaches a biological goal first (progresses most rapidly) and resembles it most closely (least selection, (9), evolutionary potential can be summarized as (accuracy/time), maximized by the best path. Assessing evolutionary plausibility as (number of accurate codes/time to evolve) implements the common-sense principle that biology implements the most accurate solution as soon as it is available. (Accurate codes/time to evolve) weights objective indices of accuracy and time equally. This may not always be wholly apt, because these quantities can vary differentially in importance, depending on pathway details. However, (accuracy/time) also increases sensitivity over individual indicators by multiplying the desirable effects of accuracy and speed, combining quantitative indicators into a single quantitative index. In Fig. 3C, (numbers of misx in 5000 environments/passages to 15 assigned fn) are used in this way to compare 16 paths to 15 encoded functions.

The results are similar to accuracy (Fig. 2B) alone, extending an observation above - quick evolution and accurate evolution are not independent, but share a logic (28). The increased sensitivity of (accuracy/speed) is evident in increased range for values in Fig. 2C, over either Fig. 2A and 2B.

So: triplet-by-triplet initiation and code evolution is again slightly aided by independently evolving environmental coding tables (tab) and by code fusions (fus), acting together (Fig. 2C). But small effects aside, we could repeat 15-function **rate findings** above: RNA world code evolution, because of reliance on pre-existing RNA capabilities (13), and rejection of alternative codes (3), is only slightly sensitive to paths (Fig. 2C).

**Adding the RNPT: time to evolve 20 SGC-like encoded functions.** Fig. 3A shows passages required to encode 20 encoded functions versus the structured list of pathways in Fig. 2A. But Fig. 2 and Fig. 3 differ prominently. Instead of differing only after radical initiation differences (Fig. 2A), time to a near-complete code (Fig. 3A) now decreases in a complex way across the plot. Code-speeded division is now a significant accelerant, whether triplets are encoded one-by-one (1SGC) or employed together for code initiation (SGC_r3). Moreover, parallel environmental codes and code fusions to them now matter in every case. Longer evolutionary paths to 20 encoded functions, through the RNPT, make evolutionary time sensitive to path details.

**Adding the RNPT: accuracy of 20 encoded functions.** Accuracy varies significantly throughout Fig. 2 and Fig. 3, because triplet assignment routinely has errors (assignment is consistently 10% random, Methods). Fig. 3B shows that accuracy increases in a roughly complementary way with the complexity of the evolutionary path. With simple beginnings (1SGC), more complex evolution helps: continuous appearance of independent coding tables (tab) and code fusions (fus) tend to yield most accurate codes. However, initiating evolution with partial early stereochemical codes (SGC_r3), the opposite is true; the simplest (notab nofus) evolution is more accurate, with and without code-sensitive division.

Simplest beginnings (1SGC Pdiv) are particularly notable: accuracy (Fig. 3B), though not speed (Fig. 3A) is stimulated by parallel code initiations (tab), and both parallel codes and fusion together (fus tab). The latter is the fusion effect on accuracy; code fusions purify nascent codes (29). While accuracy benefits are slight in the RNA world era (Fig. 3A), they are more important when enlarging early assignments via the RNPT (paths # 1-4, Fig. 3B).

**Adding the RNPT: accuracy/time for 20 encoded functions.** As an example of evolutionary plausibility ≈ accuracy/time (Fig. 3C), initial stereochemical codes (SGC_r3) can be said to be ≈ 100-fold superior to evolution using single stereochemically-guided triplets (Fig. 3C, mis0 1SGC notab nofus).

Fig. 3C also reemphasizes that the evolutionary benefits of multiple environmental codes and their fusions are specific to simple code initiation (1SGC) of RNPT codes, diminished even by code-sensitive replication (mis0 1SGC DivProb). Beginning a code with more complete early stereochemistry (SGC_r3) depresses the parallel code contribution even more. In fact, with early stereochemical initiation, simplest (notab nofus) evolution is superior, SGC_r3 DivProb notab nofus being best of all.

**Accuracy and speed.** Prior findings coupling evolutionary speed and accuracy (28) are reproduced in Fig. 3D, which plots numbers of 20-function mis0 and mis1 codes in 5000 environments, against their times to 20 functions. There is an exponentially declining relation between accurate code numbers and time to evolve, with variance greater for small numbers of codes, as expected. Thus use of mis0 (5k) codes/time to 20 encoded functions) to represent plausibility captures both the general dependence of (accuracy/time) as well as its variability from all causes. Again, evolutionary accuracy and speed do not conflict: in genetic code evolution, they co-exist.

**Summary of RNP-era codes.** Fig. 3's end-of-RNPT results reflect codes that continue from the RNA era of Fig. 2. Thus by comparing Fig. 2 and 3, we can deduce differences between the RNA era and RNA+RNPT, thereby indirectly characterizing the RNPT.

Under these conditions, codes with extensive early stereochemical assignment (SGC_r3) at the end of their RNA eras have been slightly hindered by interaction with other codes (Fig. 2C), and codes finishing RNPTs have been even more decisively retarded by code-code interactions (Fig. 3C). Thus with early extensive stereochemical assignments, least selection for the SGC mandates codes developing alone in independent environments.

Singly-initiated codes (with 10% random errors) are different; they are somewhat aided by fusion with parallel codes even after the RNA world at 15 functions (Fig. 2 C). When these codes proceed to 20 encoded functions (Fig. 3C), through the RNP era, they are very distinctly aided by independent parallel code initiations and fusions to them.

Overall differences (Fig. 2C, 3C) between SGC-like codes depend mostly on differences in code accuracy (Fig. 2B, 3B). As an illustration, it is the ≈10-fold difference in accuracy of 1SGC Pdiv fus tab codes (Fig. 3B) that makes them superior to 20-encoded-function 1SGC nofus notab codes (mis0, Fig. 3C). Thus, particularly for codes built codon-by-codon from the start, accuracy-enhancing effects of fusion with parallel independent codes in the same environment are important, first in the RNA-era, and more so during the later RNPT (Fig. 2C, 3C).

**SGC-like 20 function codes are a robust result.** Code evolution can significantly resist interference by other nascent codes (3). In Fig. 4, this is explored by varying quantities of random initiation. Codes are assembled by random single-codon-addition through the RNPT (20 assigned functions) beginning with a single assignment (1SGC DivProb tab fus). The probability of random assignment, unrelated to the SGC, during addition is varied (Prandom). While Prandom has usually been set to 0.1 (Methods); here broader non-SGC assignment is explored. Even if 1 of 5 assignments has no relation to the SGC, very SGC-like codes still appear among 5000 environments that have successfully transited the RNPT. Thus SGC-like codes continue to exist when other codes with other assignment principles can fuse with nascent SGC-like codes (see **Termination of the crescendo**, below).

**Kinetics of code completion.** In Fig. 5A, codes evolve through time from single initiations (1SGC DivProb fus tab) for reasons explained in Discussion. After a lag for code assembly, near-complete codes that have SGC assignments (mis0) are detected. Such SGC-like codes peak, then decline slowly (the crescendo; (29). This kinetic pattern also appears in the larger number of codes that differ by only one assignment from the SGC (mis1; Fig 5B).

**Completion, firstly.** SGC precursors show a crescendo at 300-400 passages, when the SGC will be most readily, most probably, selected (30). Consider two implications: firstly, the rapid rise in SGC-like codes reminds us that the Fig. 5 era hosts active turnover – assignments, decay, codon captures, new code initiations and code fusions continuously modify coding tables. Thus, while near-complete codes persist in the code population (Fig. 5A, 5B), individuals vary. For example, codes cycle through the possible configurations of 20 encoded functions and mis0. Thus the coding crescendo (Fig. 5A, 5B) is highly optimized for selection; it presents differing close SGC relatives again and again (29).

**Completion, secondly.** However, time for selection is finite. Every group in Fig. 5A and Fig. 5B slowly declines from its peak. Notably, this is true of codes closest to the SGC. Thus, probability of SGC selection also declines. Possibly, time available for resolution of incompatible code assignments can be extended by evolution of cellular coding compartments (3), so that variants can stably coexist. But there is a need for exit from the crescendo, reinforced by the first point just above. The crescendo produces a varied population containing multiple similar codes. The SGC must emerge from a crowd that resembles it. The crescendo must ultimately conclude, choosing the SGC.

**End of the crescendo.** SGC emergence can be both decisive and simple, requiring no novel mechanism. Codes within microbial carriers (9) can escape the unusual crowded environment in which they have first evolved. They enter a large, empty, different world in which organisms rarely fuse, where they predominate using rapid cell division (Fig. 1). This is simplified in Fig. 6 by following relevant escapees from the environment where they underwent their RNPT.

A minimal idea makes physical escape continuous. Organisms steadily elute from the periphery of the initial code environment, which perhaps is an unusual one, enriched with meteoric organics, or geochemical precursors or both. But escape is only effective when eluted codes become capable of thriving in a broader environment. Though escape is continuous, the first successful escapees will be those suited to a free existence.

In the example (Fig. 6), effective escape first occurs at the mean time when singly-initiated codes have evolved codes with 20 encoded functions (Path #4, Fig. 3C). The simplest case is shown, focusing on the environmental distribution of mis0 codes, that is, with assignments already identical to the SGC. As expected (8), most complete (20 function mis0) codes are a small minority (Fig. 6A). This code distribution is entered as the starting point (marked escape) on the x-axis of Fig. 6B, then cell division (Fig. 1) is followed for the first 50 passages afterward (marked escape and diaspora). The effect of exponential cell division is dramatic. Escaped codes, in a time short with respect to their evolution (Fig. 3), become more complete, due to their more rapid division (Fig. 1, Methods).

**Escape is robust, readily coupling to different prior events to produce an SGC-like code.** The differential effect of faster cell division is potentially large (Fig. 1, 6B), easily increasing relative code abundance by an order of magnitude, or more. Whatever their initial minority (e.g., Fig. 6A), accurate, complete codes can become a major fraction of a diaspora population. Escape therefore relieves (Fig. 4) the previously discussed 10% limit (29) on evolving coding accuracy by allowing rare complete and accurate codes (Fig. 4) to become predominant during diaspora (Fig. 6). Though we have analyzed a particular evolutionary moment, similar escape results (Fig. 6B) occur in other scenarios.

**Escape is numerically practical.** An example diaspora environment might be a lake of $10^{12}$ cc, ultimately harboring $10^6$ cells/ml, or $10^{18}$ total cells (marked "1st lake" in Fig. 6B). When the escaping Fig. 6A population reaches $10^{18}$ cells, it is 41.4% 20-function mis0 codes, which are completely SGC-like. Most of the remainder, a majority of 57.9%, are 19 function mis0 codes, needing only one additional correct assignment to become 20 function SGC-like codes. Accessible diaspora environments therefore yield substantial populations of SGC-like codes, in which least SGC selection (9) would seemingly be simple. For example, this sort of level distribution could easily be resolved by selection for effective information exchange between populations (31), which first exist in diaspora (Fig. 7).

One might wish to compute the escaped code population more exhaustively, expanding from the mis0 codes closest to the SGC (Fig. 6). However, only the very fastest-multiplying codes can be significant (Fig. 6B), and this limits candidates to a minor group. Escapees are also not random. They are SGC-like because all used RNA world stereochemical assignments in their initial 15 assignments (Fig. 7). Thus escaping codes, including the SGC-to-be, are strongly biased toward RNA-determined assignments, which will be "frozen" (32). Moreover, environments using code fusion (Fig. 6) are expected to reject unusual coding, due to poor fitness of code fusions with inconsistent assignments (29). So Fig. 6 seems an apt first approximation, though it surely is not unique.

One might prefer smaller lakes or cell concentrations. But similar conclusions follow in a chain of successively seeded ponds, each amplifying more complete codes after code bearers reach unpopulated downstream pools. A chain-of-lakes history amplifies the impact of the diaspora with respect to first escape, but does not change the essential mechanism. Thus, escape and diaspora from an RNPT crescendo (Fig. 6) plausibly initiated the code's present era.

**Escape and diaspora impose a selection.** No special SGC qualities are assumed in Fig. 6, save for the known catalytic superiority of more varied encoded amino acids (Fig. 1, Methods, Fig. 8). However, escape and diaspora impose their own selection. For example, the answer to questions like "which 20 amino acids?" may lie in the prerequisites for a free-living microbe at escape time.

**Crescendo, escape and diaspora are an optimal pathway.** These events provide more than RNPT exit. They create a near-ideal (33) evolutionary sequence. Crescendo supplies a changing population of highly competent codes (Fig. 5), exploring options close to the SGC. Escape and diaspora continuously monitor this shifting crescendo until a properly-adapted, free-living microbe appears (Fig. 7). A varying source of codes is continuously monitored to detect <u>both</u> rapid evolution and also accuracy (Fig. 3C), or suitability of code assignments (Fig. 2, 3). Though an SGC must evolve in a highly specialized locale, it needs no new mechanism to become adapted to a larger world.

**But code unity is not quite that simple.** Code unity must precede the Last Universal Common Ancestor (LUCA; 34) because all domains of life on Earth share a mostly common SGC (1). However, code universality has likely been reinforced many times since its plausible invention in the RNPT. Hypothetically: consider the invention of DNA (35) and the genome. Was the DNA genome not similarly decisive for the organisms that achieved it (Fig. 7)? Non-hypothetically, with data: the great extinctions of the fossil record (36), if extrapolated to all Earth life, predict a very small total probability of survival for even evolutionarily recent organisms. The code of these few survivors has become ours today. Non-hypothetically, also: the Cambrian explosion (37) carried the code that achieved complex multicellularity into all Cambrian radiation descendants. Thus the SGC had a unique pre-LUCA origin, also subsequently echoed and reinforced throughout a broadening phylogeny.

## Discussion

**The scheme for the SGC.** This text offers a hypothesis about the SGC. From specific quantitative assumptions (Methods), an evolution for the SGC is computed, invoking only well-known chemical, biological and evolutionary events. This can be elaborated or rejected in favor of a better Bayesian explanation; that is, one that more completely explains present code unity (12). Thus, Discussion's priority is to define this preferred pathway, with a synopsis of it. Fig. 7 is that summary; it expands an earlier scheme (3) which has further references.

**Sources of code assignment.** This work reveals an underlying logic for complete code assignments. It matters how forming codes assign amino acids to ribonucleotriplets. RNA-world-era assignments are only slightly dependent on other parallel codes (Fig. 2C) because 15-function codes rely mostly on multiple intrinsic RNA affinities for amino acids (13, 33); even this slight dependence on multiple codes (Fig. 2A, 2B) disappears if multiple prior stereochemical initiations occur (Fig. 2C).

Given initiation by complexes of early stereochemically-determined assignments, parallel code contributions to 20-function post-RNPT codes are likewise not useful; in fact, they are evolutionarily damaging (Fig. 3C). Assignment-stimulated code (cell) division also suppresses dependence of singly-initiated RNPT codes on other code contributions (Fig. 3C).

Accordingly, singly-initiated codes have a unique status, requiring assistance from other codes in the same environment (Fig. 3C) to reach near-completion. This is notable, because single initiation is by far the most probable primordial route to the code, indeed, to all biology. Under origin conditions, the first functional cell is improbable by definition; so the most probable primordial path mandates a single founding cell line and code. Moreover, the simplest way to elaborate this initial code is by addition of single assignments to successively solve the problems of cell elaboration (Fig. 2C, 3C). One might question the idea of codes with one assignment, but single aminoacyl-RNAs may have useful activities (38, 39).

Once having assumed the first code and cell, no further assumption is required to posit that other cells and codes progressively arise alongside the first, and that they create the SGC collaboratively, fusing their independent competencies (mis0 codes, Fig. 3C, (10). Thus single initiation-and-addition (Fig. 3C), offers both continuous code utility, uninterrupted back to its origin, and a straightforward gain-of-function mechanism suited to a challenging ancient environment. In choosing it, we resolve a previous ambiguity (28) concerning whether parallel codes are useful to code evolution.

By comparison, pre-existing initial multi-assignment stereochemical codes (e.g., SGC_r3; (3), while very effective in shaping code evolution (28), seem less probable. They efficiently evolve complete codes, but by positing an intrinsically less credible initial complexity. The also place crucial evolutionary events beyond analysis. Therefore, below we assume simpler, singly-assigned-initial routes (Fig. 2, 3) to the code, which evolve coding in view of the investigator.

**Advantage is not intuitive.** Intuition is an inadequate guide to the behavior of complex systems. Consider collaboration between fusing codes , which seems to offer a clear advantage (10). And so it does, in the case of singly-initiated codes (Fig. 3C). But with extensive stereochemical code initiation, complete coding is inhibited by parallel codes and their fusions (Fig. 3C).

Enhanced cell division (Fig. 1) also at first seems promising, but is not substantially useful until escape and diaspora (Fig. 6). Instead, it only contributes to (accuracy/speed) later when multiplication becomes the primary quantity selected (Fig. 6). This confirms prior calculations (28), which found that code-dependent divisions did not speed early code evolution. One must do the sums to decide.

**These are microbes.** That earliest organisms were microbes is almost beyond argument. But this impacted the SGC: powerful selections require microbial populations (9). In Fig. 2 and Fig. 3 legends, large numbers of environments and individual codes required are plausible only if microbes host nascent codes. Diaspora's effects (Fig. 6) depend on extensive multiplication, therefore on large microbial populations. Moreover, an essential role for code fusions (Fig. 3) suggests that early codes easily merged, most workable if they are within single-compartment, easily fusible creatures; microbes. SGC evolution was necessarily a microbial event; thus, the emerging SGC fit microbial needs.

**Before the RNA world.** At left in Fig. 7: amino acids are cosmochemicals, readily existing on the primitive Earth. This begins with the celebrated Miller-Urey experiment, in which amino acids appear in sparked flasks containing gasses that emulate a primitive reducing atmosphere (40, 41). Moreover, an overlapping set of amino acids is among the most readily synthesized in primitive oceans (42). In addition, many are in the 'cyanosulfidic set' (43) readily synthesized from plausible prebiotic chemicals (44).

Nucleotides present a more difficult synthesis, but notable progress has been made on geochemical synthesis of the pyrimidine ribonucleotides (45), and purine bases are well-known in meteoroid samples (46).

These molecules must be activated to make net synthesis of polymers plausible. Modern phosphate activation is implausible (47), but nucleotide activation as imidazolides (48) has produced very productive substitutes (49). Such activated nucleotides polymerize to make templated RNA of significant lengths (50), and longer arbitrary-sequence RNAs are made within the mineral galleries of washed, ubiquitous montmorillonite clays (51).

Early events must take place in an uncontrolled environment, where chemical contributions were undependable and sporadic. However, reactants random in amount and time are not unsurmountable obstacles (33).

**RNA world.** We suppose that short RNAs can be synthesized and replicated with moderate accuracy. Under such conditions, short peptides might be encoded by direct templating of activated, bound amino acids on RNA (4, 5). Moreover, the simplest catalytic RNAs, free 5'-5' ribodinucleotides, can be synthesized (19), perhaps even templated (52). Such cofactors readily initiate transcribed RNAs (20), perhaps extending their catalysis.

A recurring finding pertains to the capacity of RNA to specifically bind amino acids to ribonucleotide sites (13), and therefore to potentially encode those bound amino acids. Amino acid binding sites selected from large randomized RNA libraries localize 6 of 8 tested amino acids within oligoribonucleotide sequences that include coding triplets as essential functional elements. This extrapolates to 15 amino acids in a complete SGC (3). Thus the history of the code appears shaped by RNA acting alone to encode up to 15 amino acids, enough to construct peptides whose activity can exceed that of primitive RNAs (Fig. 1).

This is sufficient to synthesize catalytic peptides, e.g., to undertake complex amino acid biosyntheses. Thus the RNA world's intrinsic coding capacity is likely enough to encode peptide catalysts required to attain a modern variety of encoded amino acids. This is particularly so because ribozyme synthesis of cofactors has been demonstrated (19). Notably, the addition of simple cofactors, even NAD alone, would be sufficient to construct biosynthetic pathways to most of the current set of 20 amino acids (17, 18). A primitive RNA world translation system could construct its successor, using RNA with peptides ultimately composed of 20 amino acids, to make a ribonucleopeptide translation catalyst, the first ribosome.

Moreover, current work (Fig. 2) shows that this can occur using new single assignments. Other codes and acquisition by fusion are slightly stimulatory, but not required because evolution to 15 encoded functions occurs almost as well in their absence (Fig. 2C).

**Crescendo and RNPT.** Thus the 15 assignment capacity of RNA world translation enables crescendo and RNPT, and transition to 20 encoded amino acids. This is a form of the era previously envisioned (53, 54) for evolution via novel amino acid biosynthesis, supplementing chemically available amino acids.

**Wobble and specific translation initiation and stop coding.** Even simple accurate wobble requires sophisticated ribosomal RNA (55) and tRNA structures (56, 57). Moreover, wobble and other increases in the size of elementary coding assignments, specifically obstruct formation of immediate precursors of the SGC (11), and therefore obstruct SGC evolution. Moreover, the two last-evolving coding functions are assigned increasingly slowly for inevitable kinetic reasons (10), rationalizing later assignment by a second mechanism. For these reasons, wobble likely evolved only after other assignments had been made (Fig. 7). Start and stop codons resemble each other throughout Earth's biota, so their triplets were likely committed early, to provide 22 encoded SGC functions. But definitive start and stop biochemistries diverge in life's domains (27), and therefore likely evolved only after domain divergence (Fig. 7).

One might worry that wobble, which makes many new assignments, would be difficult to implement if begun late, with many preexisting encodings. In fact, enlargement of individually assigned regions is a general obstacle to code formation (10). But this is not a problem here: the subset of SGC-like codes has necessarily placed its assignments so they can be extended by wobble without substantially disrupting its SGC resemblance. Late wobble assignments have been discussed (10), alongside graphics illustrating their practicality and their harmony with other chemical order in the SGC.

**Escape and diaspora.** Evolution would first yield codes with a distribution of capabilities, of which a minority would be immediate precursors of the SGC (Fig. 6A). The favored minority have, by chance, taken a simpler path to the final SGC (8), the path of least selection (9). Thus, a mixture of codes must be resolved (Fig. 5) to find a minority SGC. Here it is suggested that the translational proficiency of the full code (Fig. 1), elaborated through the RNPT, is the key to SGC resolution. The entire distribution of codes can escape from their special origin site (Fig. 6), but faster cell division by SGC-containing microbes ensures that they become abundant during diaspora. Escape and diaspora are shown at a particular time (Fig. 7) at which they become important, but these events were probably continuous. Pervasive escape is not only simpler, but also selects for fast evolution (Fig. 3). This straightforward pathway provides varied codes, then values competent free cell division (Methods, Fig. 1, 8). It therefore ensures that the SGC exists for the LUCA, and subsequently, in all domains of life on Earth.

**Biology as anthology.** This maxim (11) refers to Biology's distinctive ability to gather information derived in separate loci or organisms into a more capable whole. Anthology was first defined to characterize the fusion of codes to rapidly yield a more complete SGC (10). But here we extend its definition: anthological advantage includes the significantly more capable cell which results from joining multiple innovations. This fixes the advantage gained by joining, and can make multiple new advantages universal, as for the SGC and its effect on cell multiplication (Fig. 6).

Possessors of any complex new property will have solved a demanding evolutionary problem by taking a less likely, but more favorable, evolutionary path. Complex, inevitably improbable, constructs like the SGC appear as minorities (Fig. 6A) in a large population (10). They usually require rescue from their minority status. The more exceptional their new capacity, the smaller the likely number of innovators. Escape and diaspora, interpreted broadly, manifest the superiority of newly-evolved capacities, implementing this rescue. Therefore anthology, innovation by gathering, is a likely source of evolutionary progress, on Earth and elsewhere.

**Methods**

**Code evolution.** Initial code development is simulated in multiple environments that can contain any number of evolving codes (8). Time is measured in passages, which are short fixed times during which only one evolutionary event will occur. Computation can be stopped in varied ways; for example, when the first code with 20 encoded functions evolves in an environment. For present data, results are governed by:

```
 ini_nr= 1          Number of initial assignments; if 0.0, SGC_r3 initiates evolution
 Pra_ini= 0.0       Probability that initial assignments are random, non-SGC
 tab_nr= 1          Number of triplet assignments made after initial initiation
 Pra_tab= 0.2       Probability of random non-SGC assignments, 0.0 to 0.2, usually 0.1
 new_nr= 1          Number of initial triplets assigned in later parallel code
 Pra_new= 0.0       Probability of initial random assignments in later parallel code
 Pmut= 0.00975      Probability of capture of a triplet a single mutation away, for a current or related aa
 Pdec= 0.00975      Probability of decay of an existing assignment, leaving its triplet unassigned
 Pinit= 0.075       Probability of initiation of a coding table
 Pfus= 0.000125     Probability of code fusion per available partner: set to 0.0 (nofus) or 0.000125 (fus)
 Ptab= 0.08         Probability of initiation of later coding tables: set to 0.0 (notab) or 0.08 (tab)
 Pwob= 0.0          Probability of initiation of simple wobble coding; set @ 0.0 for late wobble
 Pdivn= 0.046       Probability of code division, 0.046 to 0.958 for 1 to 20 amino acids encoded (Fig.1)
```

Probabilities are for the event during a single passage.

Code used for these calculations, with internal modifications commented for different models, has been deposited as Pascal text Ct25d.pas at Zenodo: 10.5281/Zenodo 11457391. Code was written, compiled and executed in console mode, using the Lazarus developmental environment 2.2.ORC1 r65149 with Free Pascal 3.2.2 as compiler, running under Microsoft Windows 10. Tab-delimited results were passed to Micosoft Excel 2016 for analysis and graphing.

**Code division.** It is assumed that physical cell division requires 0.05 the time for biosynthesis and division under control of RNA catalysis. Specific activities for a nucleotide diphosphate kinase with varied numbers of amino acids were taken from (15) and (14), some being corrected to 50°C using published activation studies (Fig. 8). The dotted least squares line suggests a free energy contribution of ≈ - 0.8 kCal/mol to accelerate catalysis per amino acid, but also that different amino acids vary considerably, with contributions varying ≈ ± 10-fold from the conceptual average. These data were combined with the estimate that proteins are about 1000-fold more capable than ribozymes in the same reaction (25) to make Fig. 1. Fig. 1 begins (at 10 encoded functions) with RNA catalyzed cell and code division, then takes account of Fig. 8 data on mean increased rates for diphosphate kinase. At 20 functions encoded, there is faster preparative biosynthesis, but division is still rate-limited, now by distribution of cell components.

**References**

1. Y. Shulgina, S. R. Eddy, A computational screen for alternative genetic codes in over 250,000 genomes. *eLife* **10**, e71402 (2021).

2. L. Ribas de Pouplana, The evolution of aminoacyl-tRNA synthetases: From dawn to LUCA. *The Enzymes* **48**, 11–37 (2020).

3. M. Yarus, From initial RNA encoding to the Standard Genetic Code. [Preprint] (2023). Available at: https://www.biorxiv.org/content/10.1101/2023.11.07.566042v2 [Accessed 12 December 2023].

4. M. Yarus, Amino Acids as RNA Ligands: a Direct-RNA-Template Theory for the Code's Origin. *J Mol Evol* **47**, 109–117 (1998).

5. R. M. Turk-Macleod, D. Puthenvedu, I. Majerfeld, M. Yarus, The plausibility of RNA-templated peptides: simultaneous RNA affinity for adjacent peptide side chains. *J Mol Evol* **74**, 217–25 (2012).

6. G. Q. Tang, H. Hu, J. Douglas, C. W. Carter Jr, Primordial aminoacyl-tRNA synthetases preferred minihelices to full-length tRNA. *Nucleic Acids Res.* gkae417 (2024). https://doi.org/10.1093/nar/gkae417.

7. J. C. Bowman, A. S. Petrov, M. Frenkel-Pinter, P. I. Penev, L. D. Williams, Root of the Tree: The Significance, Evolution, and Origins of the Ribosome. *Chem. Rev.* **120**, 4848–4878 (2020).

8. M. Yarus, Evolution of the Standard Genetic Code. *J. Mol. Evol.* **89**, 19–44 (2021).

9. M. Yarus, Evolution and favored change: a principle of least selection. [Preprint] (2024). Available at: https://www.biorxiv.org/content/10.1101/2021.06.27.450095v4 [Accessed 28 May 2024].

10. M. Yarus, Fitting the standard genetic code into its triplet table. *Proc. Natl. Acad. Sci. U. S. A.* **118**, e2021103118 (2021).

11. M. Yarus, Ordering events in a developing genetic code. *RNA Biol.* **21**, 1–8 (2024).

12. M. Yarus, J. G. Caporaso, R. Knight, Origins of the genetic code: the escaped triplet theory. *Annu Rev Biochem* **74**, 179–98 (2005).

13. M. Yarus, The Genetic Code and RNA-Amino Acid Affinities. *Life* **7**, 13 (2017).

14. M. Kimura, S. Akanuma, Reconstruction and Characterization of Thermally Stable and Catalytically Active Proteins Comprising an Alphabet of ~ 13 Amino Acids. *J. Mol. Evol.* **88**, 372–381 (2020).

15. R. Shibue, *et al.*, Comprehensive reduction of amino acid set in a protein suggests the importance of prebiotic amino acids for stable proteins. *Sci. Rep.* **8**, 1227 (2018).

16. H. B. White, Coenzymes as Fossils of an Earlier Metabolic State. *J Mol Evol* **7**, 101–104 (1976).

17. A. Kirschning, Coenzymes and Their Role in the Evolution of Life. *Angew. Chem. Int. Ed Engl.* **60**, 6242–6269 (2021).

18. A. Kirschning, On the Evolutionary History of the Twenty Encoded Amino Acids. *Chem. Weinh. Bergstr. Ger.* **28**, e202201419 (2022).

19. F. Huang, C. W. Bugg, M. Yarus, RNA-Catalyzed CoA, NAD, and FAD synthesis from phosphopantetheine, NMN, and FMN. *Biochemistry* **39**, 15548–55 (2000).

20. F. Huang, Efficient incorporation of CoA, NAD and FAD into RNA by in vitro transcription. *Nucleic Acids Res.* **31**, e8 (2003).

21. C. W. Carter, What RNA World? Why a Peptide/RNA Partnership Merits Renewed Experimental Attention. *Life Basel Switz.* **5**, 294–320 (2015).

22. K. Kruger, *et al.*, Self-Splicing RNA: Autoexcision and Autocyclization of the Ribosomal RNA Intervening Sequence of Tetrahymena. *Cell* **31**, 147–157 (1982).

23. C. Guerrier-Takada, K. Gardiner, T. Marsh, N. Pace, S. Altman, The RNA Moiety of Ribonuclease P Is the Catalytic Subunit of the Enzyme. *Cell* **35**, 849–857 (1983).

24. T. J. Wilson, D. M. J. Lilley, The potential versatility of RNA catalysis. *Wiley Interdiscip. Rev. RNA* **12**, e1651 (2021).

25. G. J. Narlikar, D. Herschlag, Mechanistic aspects of enzymatic catalysis: lessons from comparison of RNA and protein enzymes. *Annu. Rev. Biochem.* **66**, 19–59 (1997).

26. W. Gilbert, The RNA world. *Nature* **319**, 618 (1986).

27. A. M. Burroughs, L. Aravind, The Origin and Evolution of Release Factors: Implications for Translation Termination, Ribosome Rescue, and Quality Control Pathways. *Int. J. Mol. Sci.* **20**, 1–24 (2019).

28. M. Yarus, The Genetic Code Assembles via Division and Fusion, Basic Cellular Events. *Life Basel Switz.* **13**, 2069 (2023).

29. M. Yarus, A crescendo of competent coding (c3) contains the Standard Genetic Code. *RNA N. Y. N* rna.079275.122 (2022). https://doi.org/10.1261/rna.079275.122.

30. M. Yarus, Optimal Evolution of the Standard Genetic Code. *J. Mol. Evol.* **89**, 45–49 (2021).

31. K. Vetsigian, C. Woese, N. Goldenfeld, Collective evolution and the genetic code. *Proc Natl Acad Sci USA* **103**, 10696–701 (2006).

32. F. H. C. Crick, The Origin of the Genetic Code. *J Mol Biol* **38**, 367–379 (1968).

33. M. Yarus, A ribonucleotide origin for life - fluctuation and near-ideal reactions. *Orig Life Evol Biosph* **43**, 19–30 (2013).

34. W.-K. Mat, H. Xue, J. T.-F. Wong, The genomics of LUCA. *Front. Biosci. J. Virtual Libr.* **13**, 5605–5613 (2008).

35. J. P. Dworkin, A. Lazcano, S. L. Miller, The roads to and from the RNA world. *J. Theor. Biol.* **222**, 127–134 (2003).

36. P. Hull, Life in the Aftermath of Mass Extinctions. *Curr. Biol. CB* **25**, R941-952 (2015).

37. S. A. Bowring, *et al.*, Calibrating Rates of Early Cambrian Evolution. *Science* **261**, 1293–1298 (1993).

38. E. Szathmáry, The origin of the genetic code: amino acids as cofactors in an RNA world. *Trends Genet* **15**, 223–9 (1999).

39. A. Radakovic, S. DasGupta, T. H. Wright, H. R. M. Aitken, J. W. Szostak, Nonenzymatic assembly of active chimeric ribozymes from aminoacylated RNA oligonucleotides. *Proc. Natl. Acad. Sci. U. S. A.* **119**, e2116840119 (2022).

40. S. L. Miller, Production of amino acids under possible primitive earth conditions. *Science* **117**, 528–529 (1953).

41. J. L. Bada, New insights into prebiotic chemistry from Stanley Miller's spark discharge experiments. *Chem. Soc. Rev.* **42**, 2186–2196 (2013).

42. P. G. Higgs, R. E. Pudritz, A thermodynamic basis for prebiotic amino acid synthesis and the nature of the first genetic code. *Astrobiology* **9**, 483–490 (2009).

43. B. H. Patel, C. Percivalle, D. J. Ritson, Colm. D. Duffy, J. D. Sutherland, Common origins of RNA, protein and lipid precursors in a cyanosulfidic protometabolism. *Nat. Chem.* **7**, 301–307 (2015).

44. J. D. Sutherland, The Origin of Life—Out of the Blue. *Angew. Chem. Int. Ed.* **55**, 104–121 (2016).

45. M. W. Powner, B. Gerland, J. D. Sutherland, Synthesis of activated pyrimidine ribonucleotides in prebiotically plausible conditions. *Nature* **459**, 239–42 (2009).

46. M. P. Callahan, *et al.*, Carbonaceous meteorites contain a wide range of extraterrestrial nucleobases. *Proc. Natl. Acad. Sci.* **108**, 13995–13998 (2011).

47. F. H. Westheimer, Why nature chose phosphates. *Science* **235**, 1173–8 (1987).

48. R. Lohrmann, L. E. Orgel, Prebiotic Activation Processes. *Nature* **244**, 418 (1973).

49. N. Prywes, J. C. Blain, F. D. Frate, J. W. Szostak, Nonenzymatic copying of RNA templates containing all four letters is catalyzed by activated oligonucleotides. *eLife* **5**, e17756 (2016).

50. P. K. Bridson, L. E. Orgel, Catalysis of accurate poly(C)-directed synthesis of 3'-5'-linked oligoguanylates by Zn2+. *J Mol Biol* **144**, 567–77 (1980).

51. J. P. Ferris, A. R. Hill, R. Liu, L. E. Orgel, Synthesis of long prebiotic oligomers on mineral surfaces. *Nature* **381**, 59–61 (1996).

52. M. Yarus, Eighty routes to a ribonucleotide world; dispersion and stringency in the decisive selection. *RNA N. Y. N* **24**, 1041–1055 (2018).

53. J. T.-F. Wong, A Co-Evolution Theory of the Genetic Code. *Proc Natl Acad Sci USA* **72**, 1909–1912 (1975).

54. J. T. Wong, Evolution of the genetic code. *Microbiol Sci* **5**, 174–81 (1988).

55. J. M. Ogle, *et al.*, Recognition of Cognate Transfer RNA by the 30S Ribosomal Subunit. *Science* **292**, 897–902 (2001).

56. D. W. Schultz, M. Yarus, tRNA structure and ribosomal function. I. tRNA nucleotide 27-43 mutations enhance first position wobble. *J Mol Biol* **235**, 1381–94 (1994).

57. S. Ledoux, M. Olejniczak, O. C. Uhlenbeck, A sequence element that tunes Escherichia coli tRNA(Ala)(GGC) to ensure accurate decoding. *Nat. Struct. Mol. Biol.* **16**, 359–364 (2009).

**Figure legends.**

**Figure 1.** Probability of code division (Pdiv) during a passage, related to the number of functions in the code of an organism. As the capabilities of its catalysts increase due to an increasing variety of residues in its peptides, the rate at which an organism divides can increase (Methods).

**Figure 2A.** Mean time in passages to evolve a code with 15 encoded functions, using different evolutionary pathways. These data are means from ≈ 5100 environments for each path; environments evolved a mean of 1.9 to 10.7 internal codes to reach 15 functions encoded. Pathways are numbered 1 to 16 on the abscissa, covering all possible combinations of path options. The upper figure legend defines paths 1 – 16, using the following abbreviations: **1SGC** = coding initiates with one SGC assignment; **SGC_r3** = coding initiates with 14 SGC assignments, the extrapolated number due to all SGC oligonucleotide: amino acid interaction. **Pdiv** = cell and code division is constant, at the rate attributable to RNA catalysis alone. **DivProb** = cell and code division occurs at the rate in Fig. 1, allowing for acceleration due to peptide catalysis. **notab** = the initial code and its descendants are the only codes in their environment. **tab** = new independent codes arise (with one assignment) at the same rate as the initial code. **nofus** = codes do not fuse. **fus** = codes from any origin can fuse with another live code. If assignments are compatible, a new fused code with all assignments exists. If fused code assignments conflict (a codon has >1 function), fused codes go extinct.

**Figure 2B.** Accuracies for codes that have evolved 15 encoded functions. Pathways are the same as Fig. 2A. Accuracy is measured as the number of codes per 5000 evolutionary environments (**mis0 (5k)**) that have no misassignments when compared to the SGC, or the number of codes that have one misassignment in 5000 independent environments (**mis1 (5k)**).

**Figure 2C.** Evolutionary plausibility of 16 pathways to 15 encoded functions. Plausibility is estimated as accuracy divided by time to evolve, thereby combining evolutionary time (Fig. 2A) and accuracy (Fig. 2B) into a single sensitive index consistent with a principle of least selection (see text: **Summary of RNA era codes: an index of plausibility**)**.**

**Figure 3A.** Mean time in passages to evolve a code with 20 encoded functions, using different evolutionary pathways. These data are means from 22,500 to 43,700 environments for each path; environments evolved a mean of 12 to 49 internal codes to reach 20 functions encoded. Pathways are numbered 1 to 16 on the abscissa, as in Fig. 2A.

**Figure 3B.** Accuracies for codes that have evolved 20 encoded functions. Pathways are the same as Fig. 2A. Accuracy is plotted as the number of codes per 5000 evolutionary environments (**mis0 (5k)**) that have no misassignments when compared to the SGC, or the number of codes that have one misassignment in 5000 independent environments (**mis1 (5k)**).

**Figure 3C.** Evolutionary plausibility of 16 pathways to 20 encoded functions. Plausibility is estimated as accuracy divided by time to evolve, thereby combining evolutionary time (Fig. 3A) and accuracy (Fig. 3B) into a single sensitive index consistent with a principle of least selection (see text: **Summary of RNA era codes: an index of plausibility**)**.**

**Figure 3D.** Evolutionary speed and accuracy are related. Data are for all 16 possible paths to 20 encoded functions; 1SGC/SGC_r3, Pdiv/DivProb, fus/nofus, tab/notab.

**Figure 4.** SGC-like codes decline, but still exist if assignments are randomized. Data are means of 15000 singly-initiated, 1SGC DivProb fus tab environments (Path #8) evolved to contain a 20 function code: this required a mean of 44 to 52 codes per environment. Numbers of mis0 and mis1 codes per 5000 environments (mis0 (5k) and mis1 (5k)) are plotted versus fraction random, singly-added triplet assignments.

**Figure 5A.** Kinetics for SGC-like codes. The number of SGC-like codes per 5000 environments (mis0 (5k)) is plotted versus time in passages, though the code crescendo, for several levels of near-complete encoding. Points are means for 5000 to 25000 1SGC DivProb fus tab (Path #8) environments.

**Figure 5B.** Kinetics for SGC-like codes. The same kinetic calculations as Fig. 5A, but mean data for a second SGC-like code, mis1 (5k) are plotted.

**Figure 6A.** Escape and diaspora increase complete coding. A plausible beginning distribution for successful escape is plotted. Initial escapees here are SGC-like (mis0) codes from 15000 1SGC DivProb nofus notab environments containing 559,000 total codes at a mean of 419 passages. There are 37.3 codes in the mean environment at the moment of successful escape.

**Figure 6B.** Fate of codes during an escape and diaspora. Codes of Fig. 6A are followed for 50 passages after escape. The number of codes is expressed in powers of 10. Cell division is assumed at the rates of Fig. 1. **1$^{st}$ lake** is the initial site of escape discussed in **Results**. **RNA** is division at the rate allowed by initial RNA catalysis. **RNPN** is division at rates for a protein enzyme having **N** encoded amino acids (Fig. 1, 8).

**Figure 7.** Graphic summary of SGC evolution. Colored areas are emphasized in this work. Time increases from left to right, but breadths are not intended to accurately reflect intervals. Events occur at the position of the initial letter of their description: for example, "SGC exists.." means that the SGC exists at the first letter "S". See **Discussion** for details and references.

**Figure 8.** Specific activities at 50°C for a nucleotide diphosphate kinase (Arc1) with the deduced sequence of an Archaeal/Bacterial ancestor, reconstructed with reduced numbers of amino acids (14, 15). The increasing activity of the least squares dotted line was used to construct Fig. 1 (Methods).

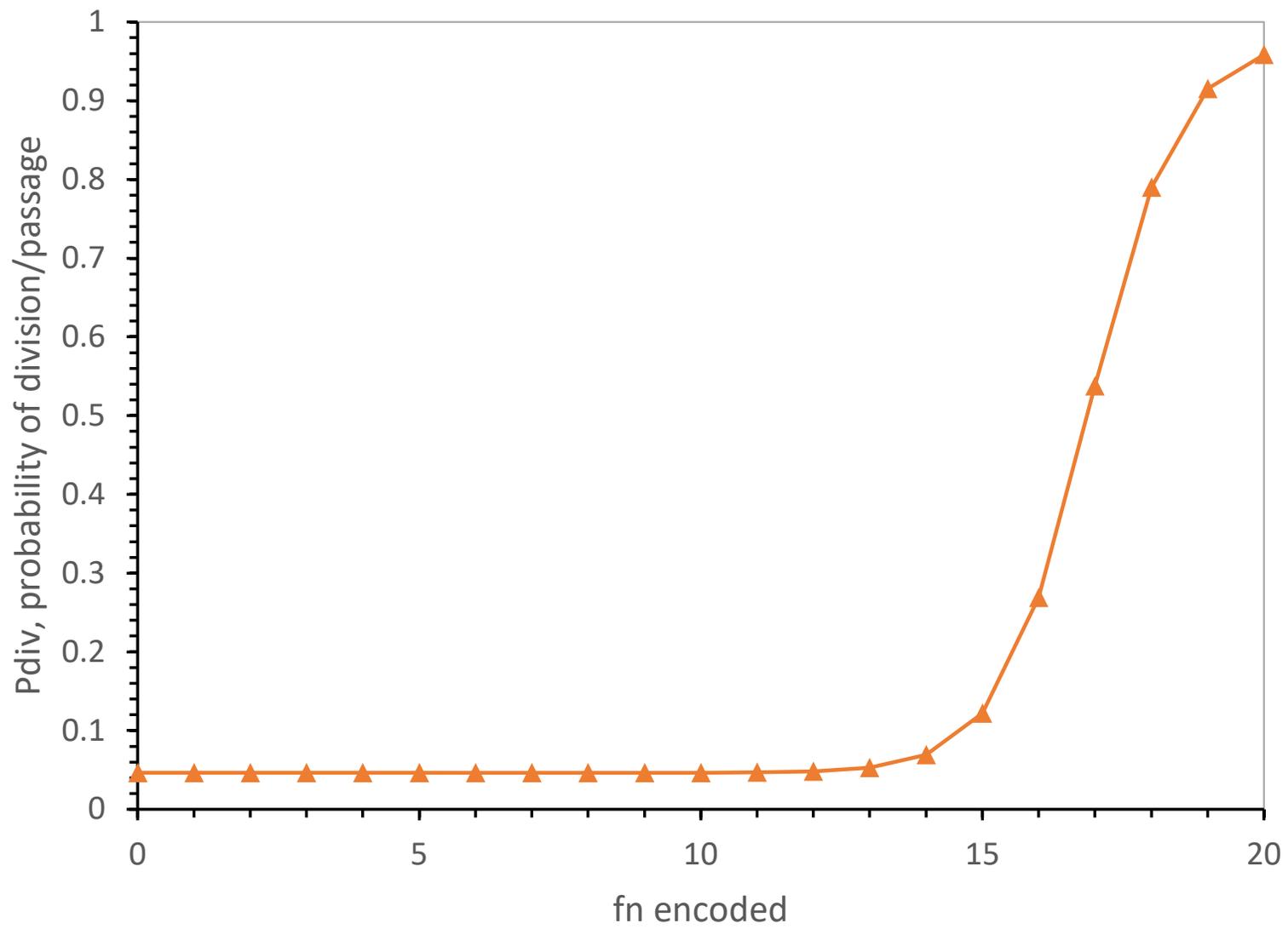

Figure 1

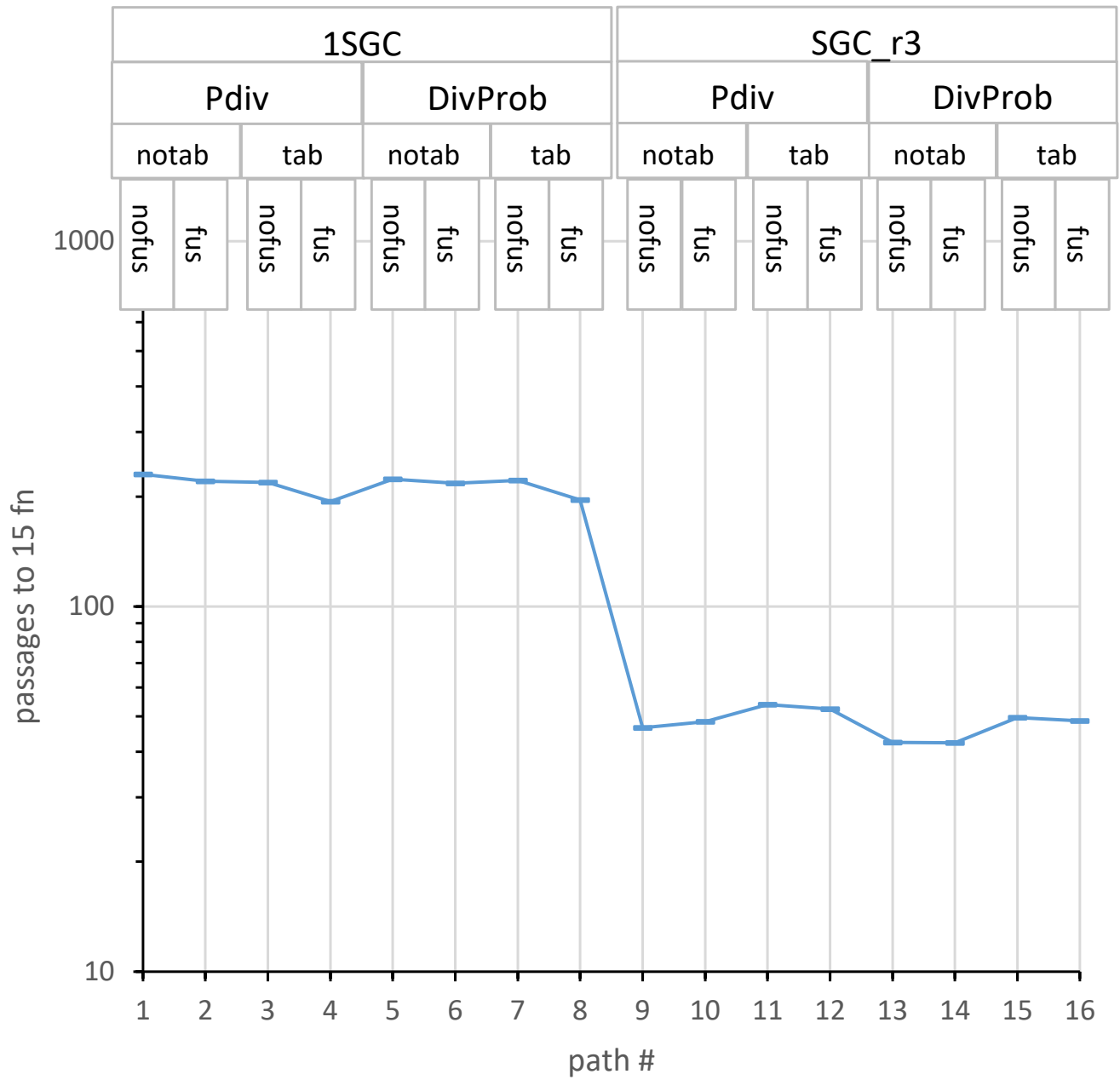

Figure 2A

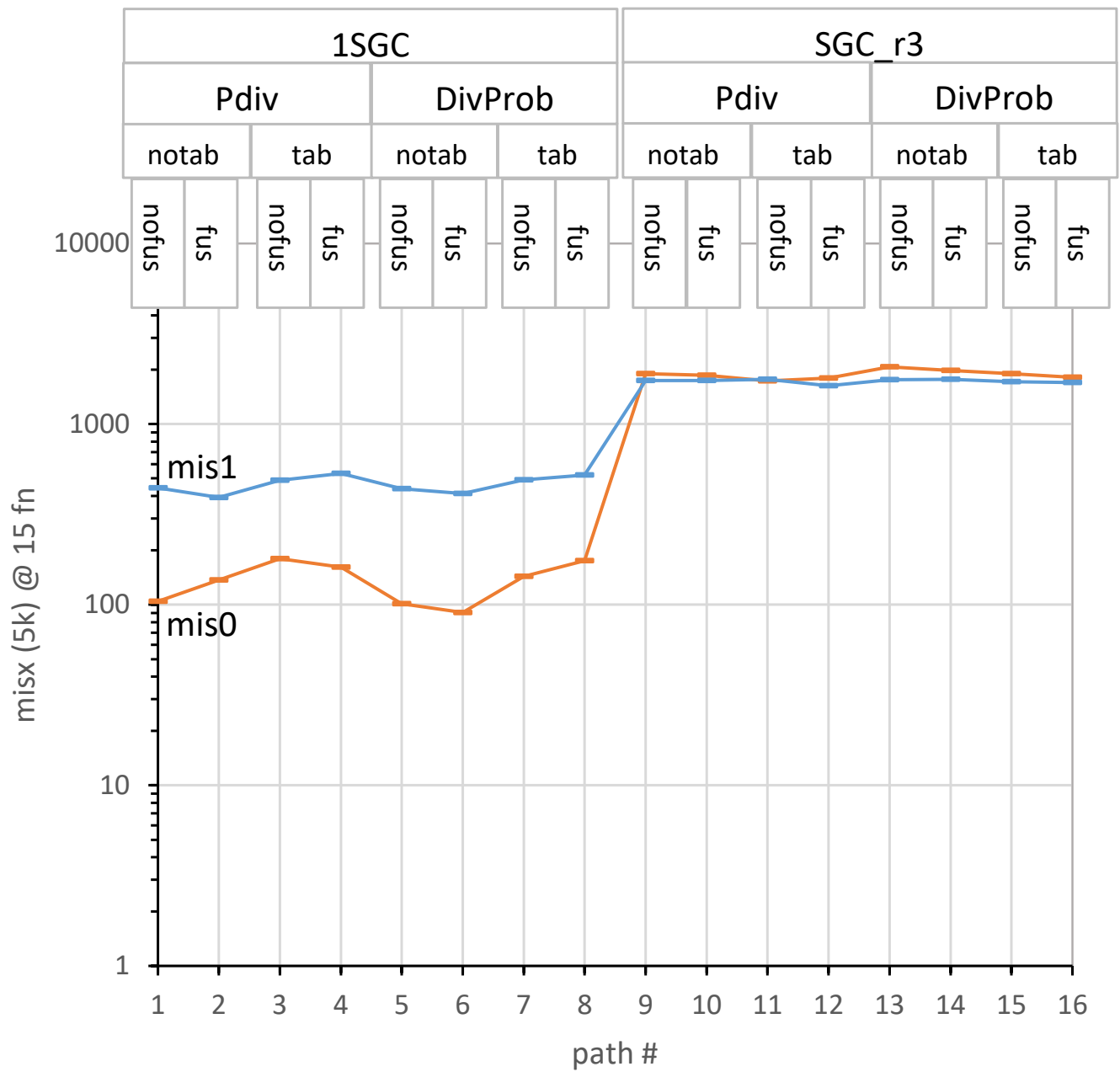

Figure 2B

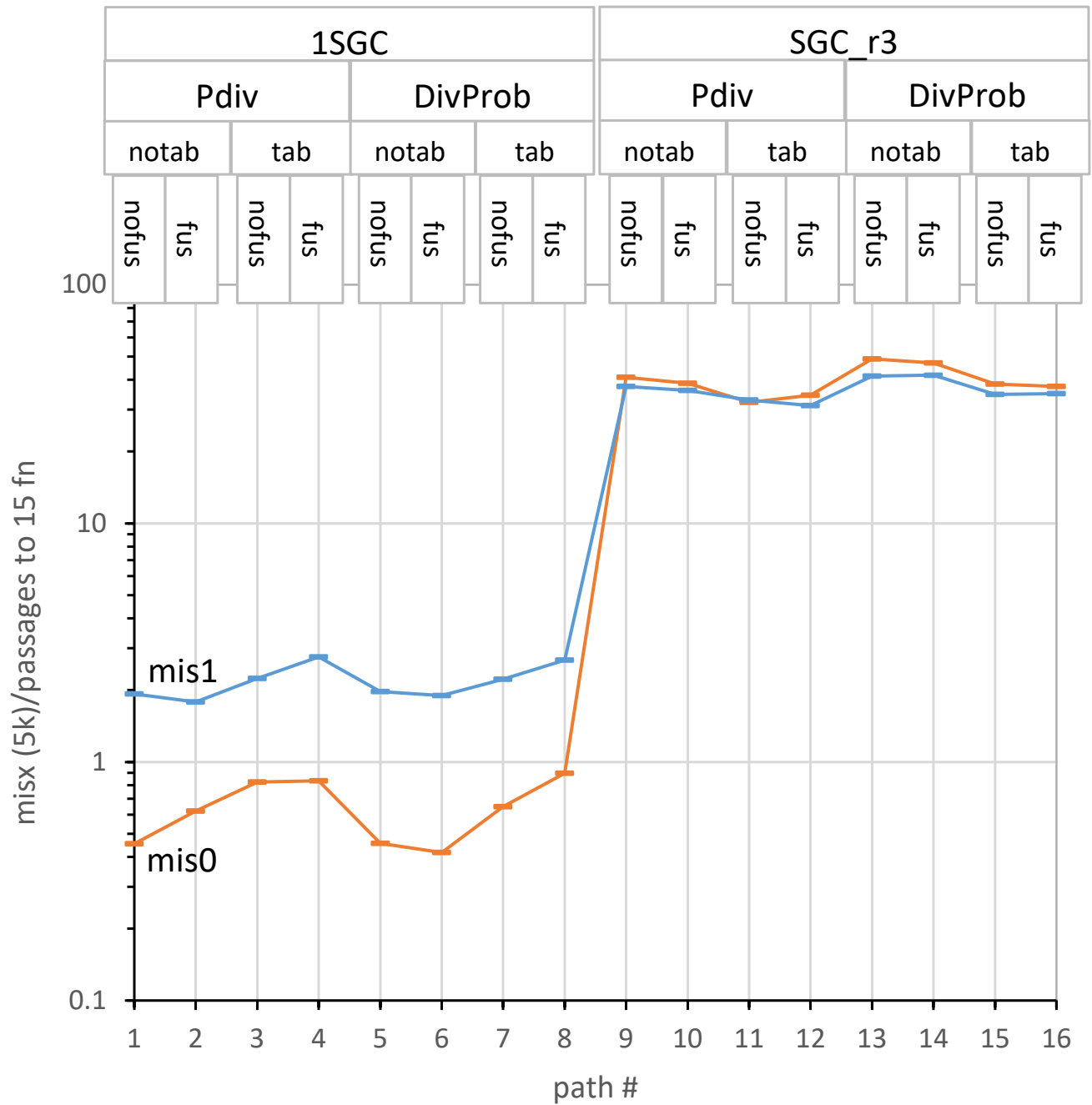

Figure 2C

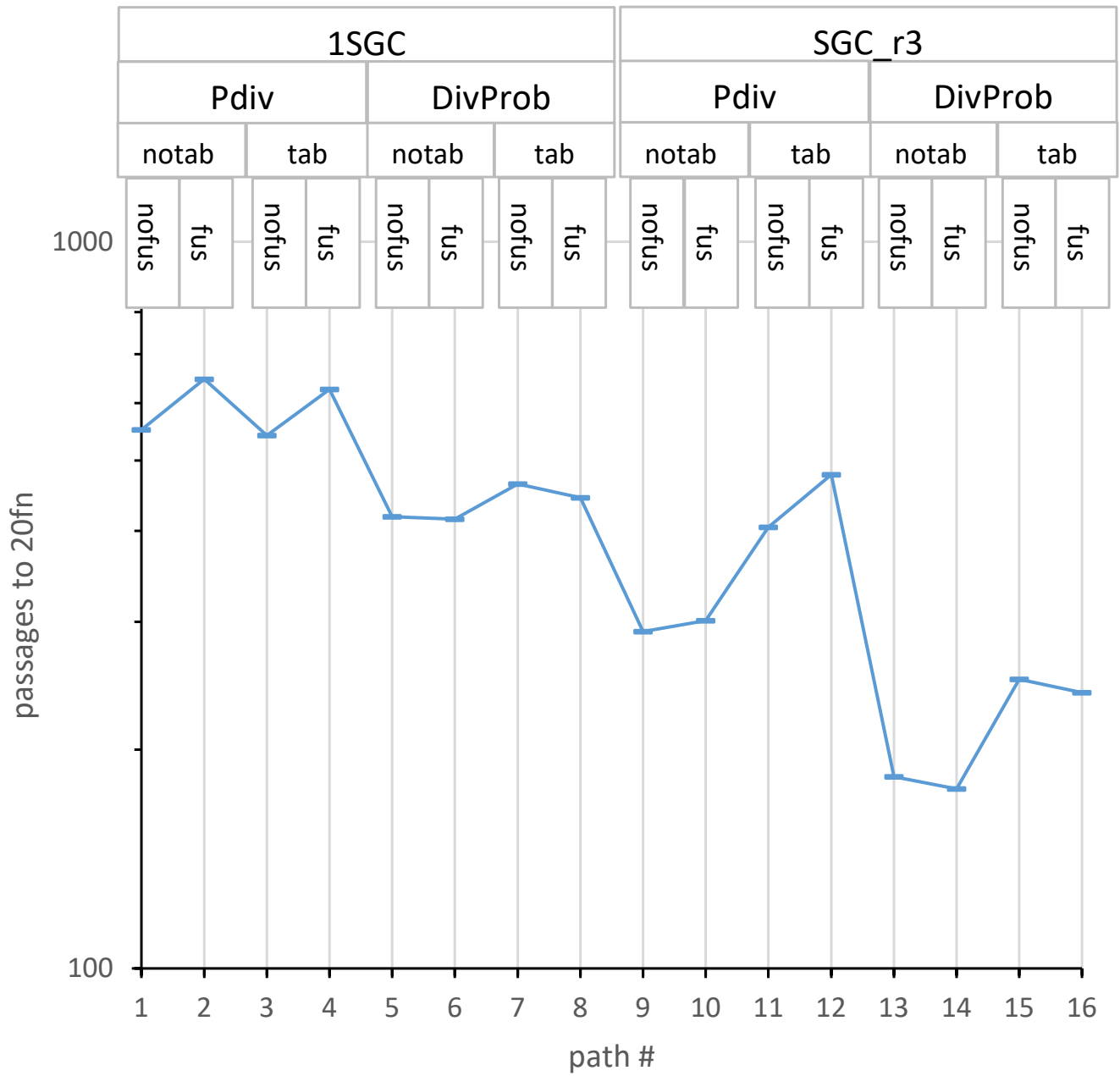

Figure 3A

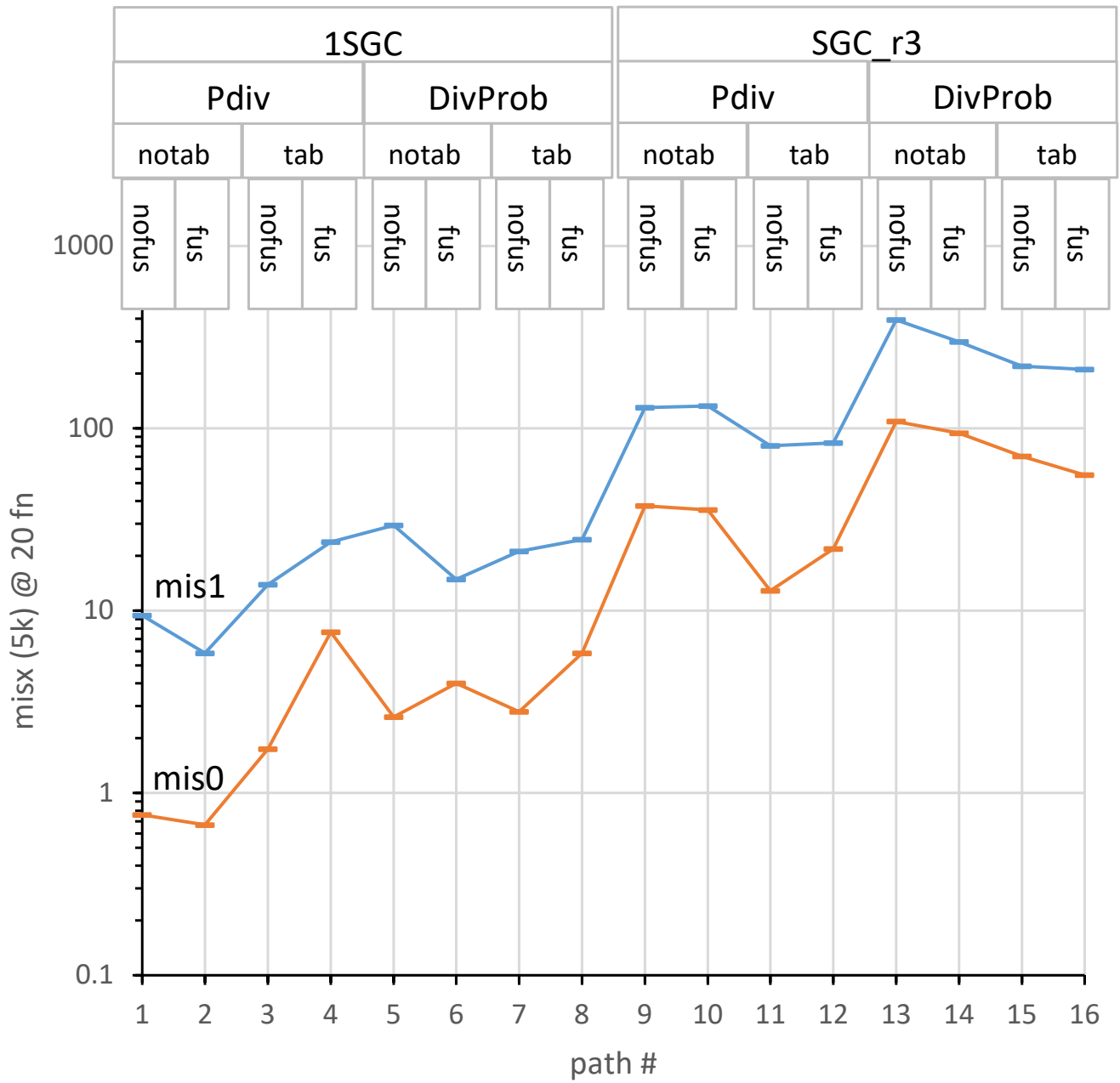

Figure 3B

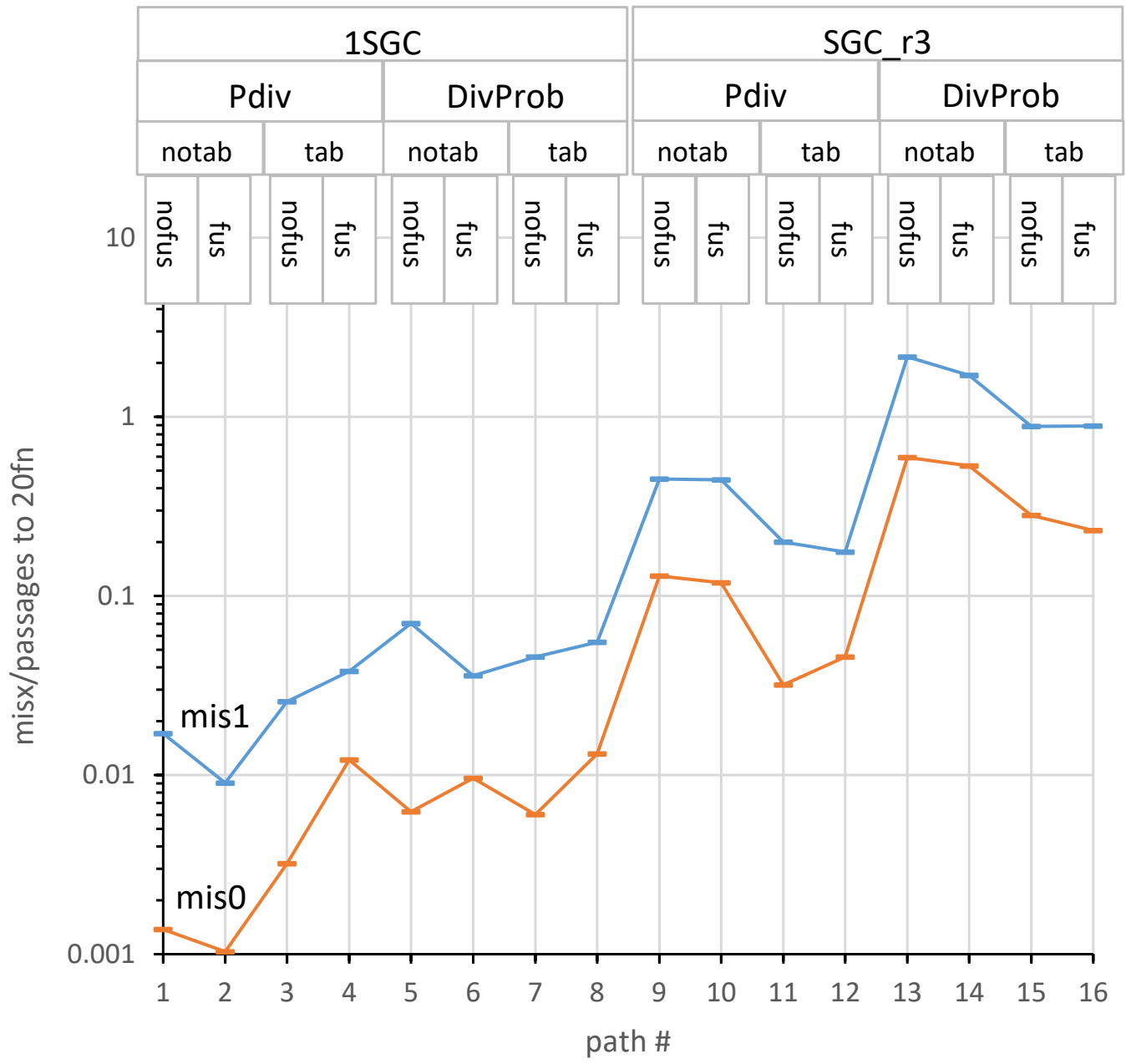

Figure 3C

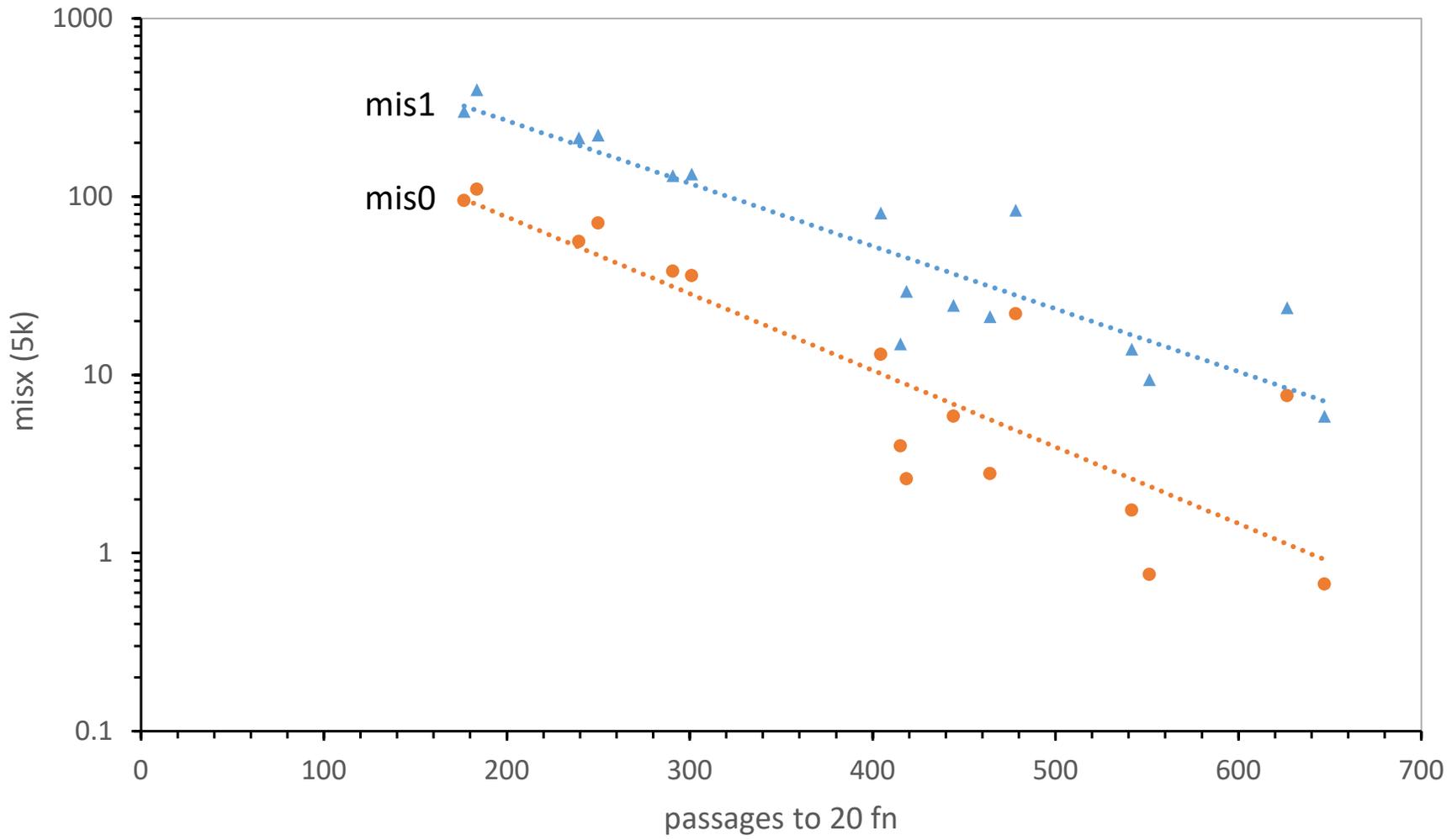

Figure 3D

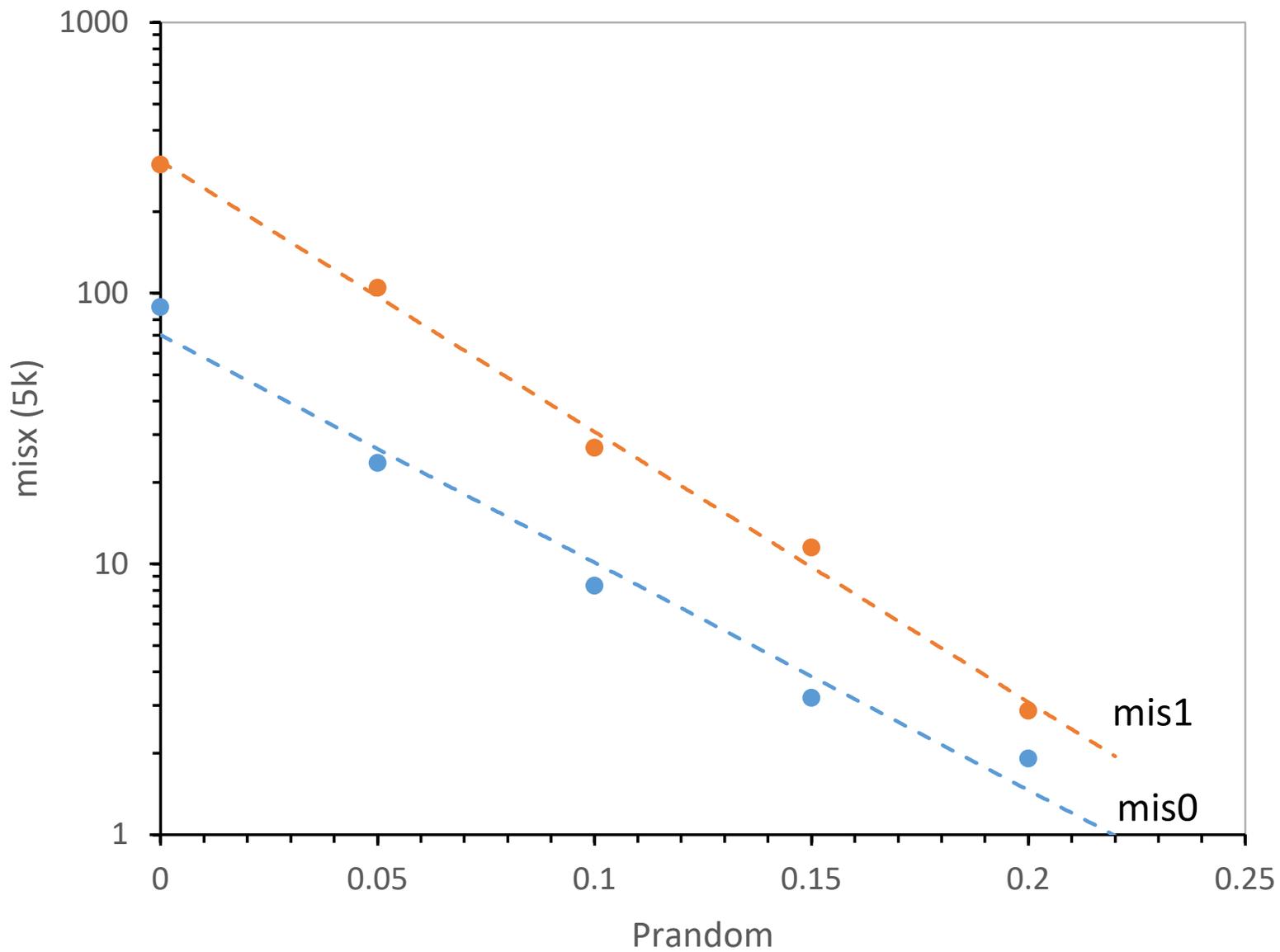

Figure 4

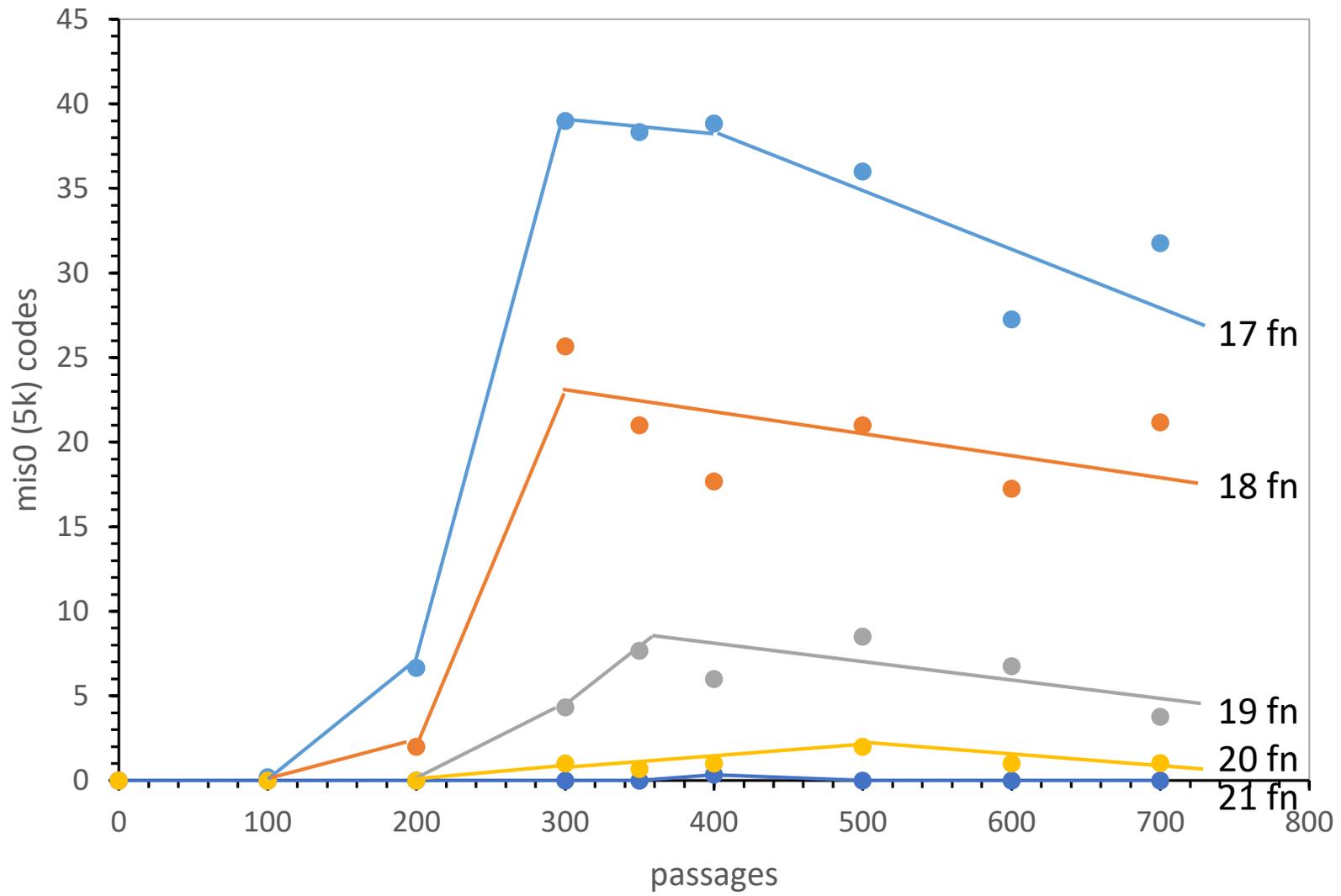

Figure 5A

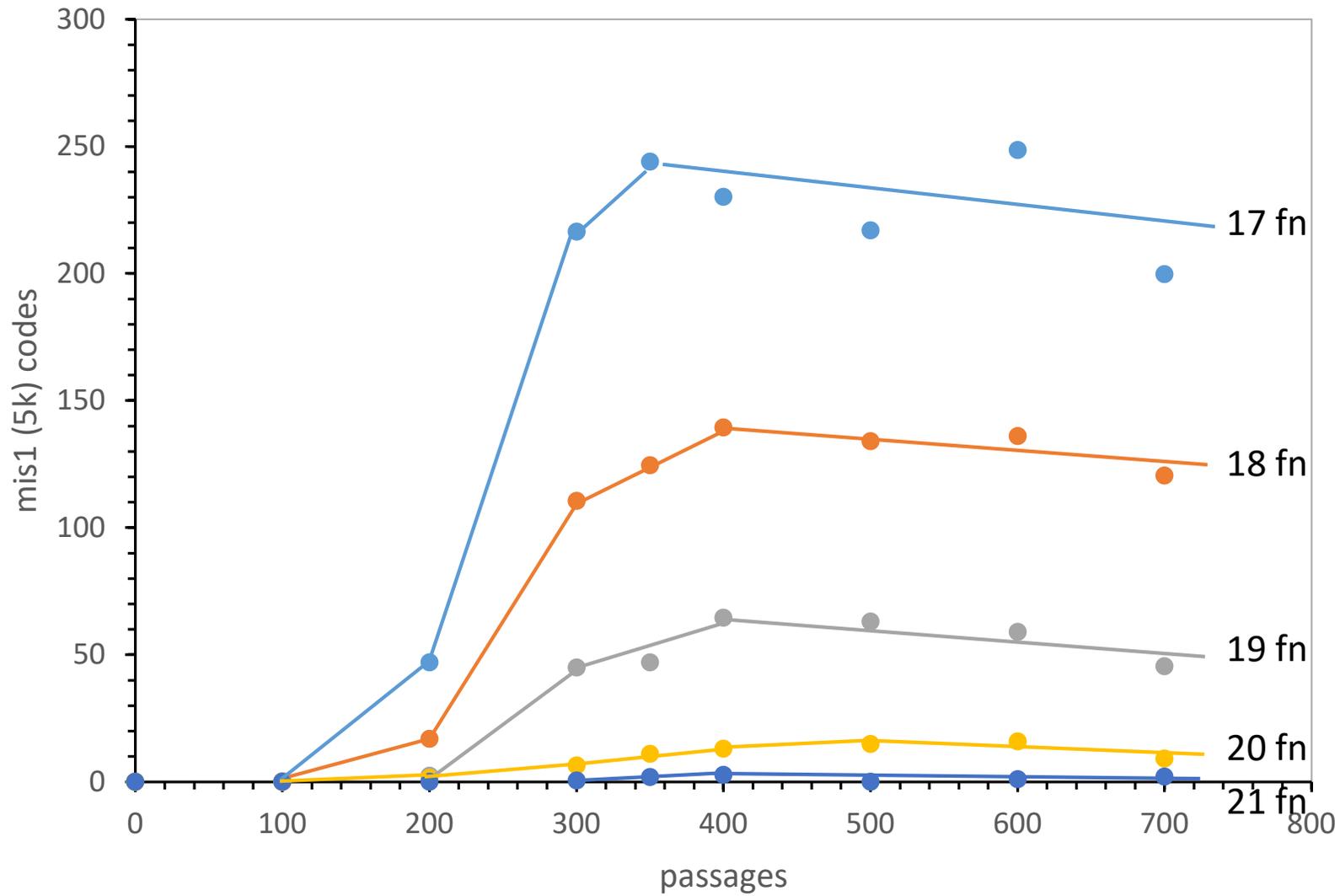

Figure 5B

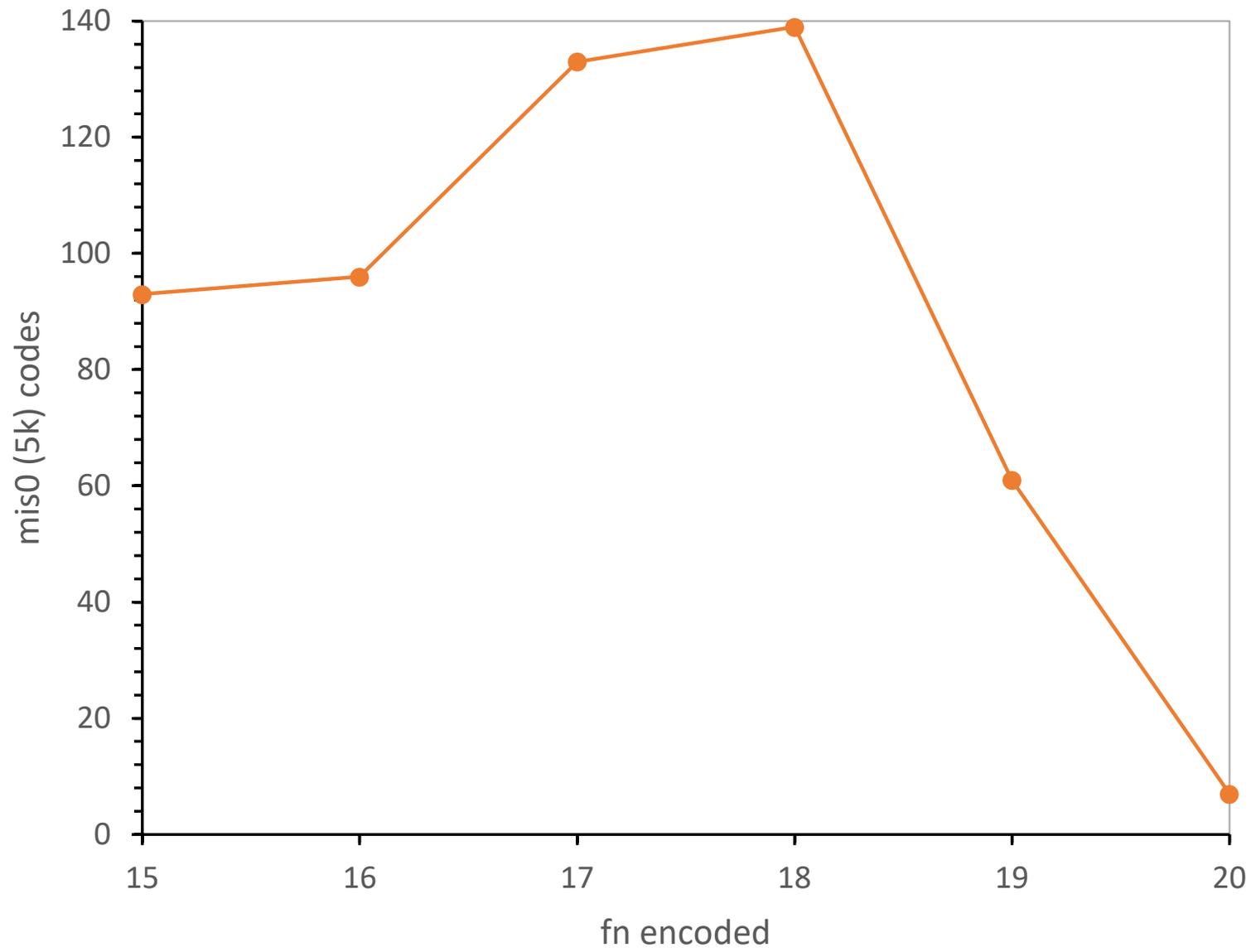

Figure 6A

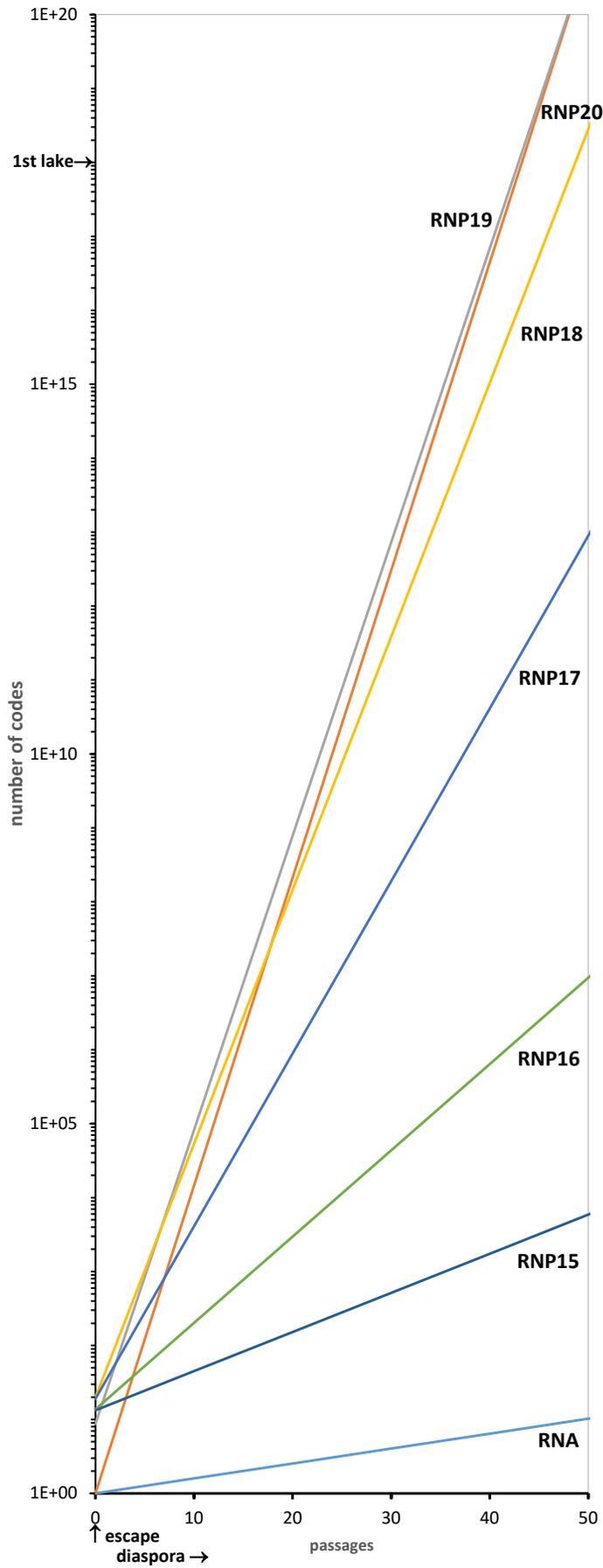

Figure 6B

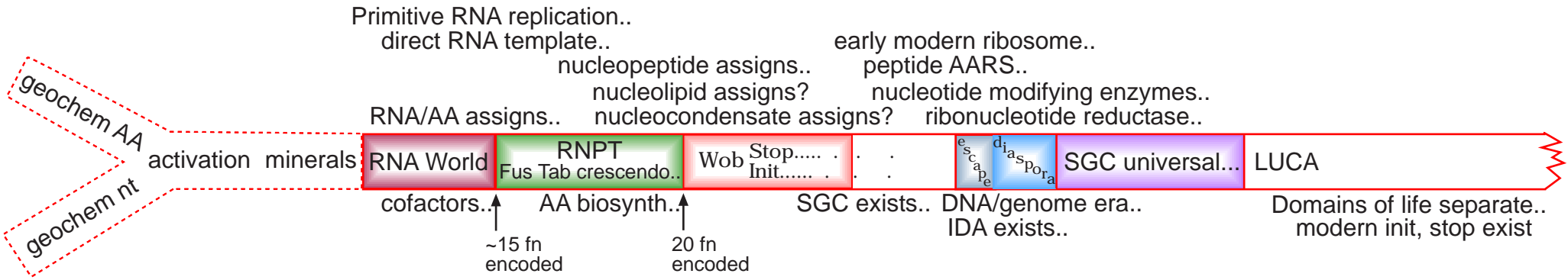

Figure 7

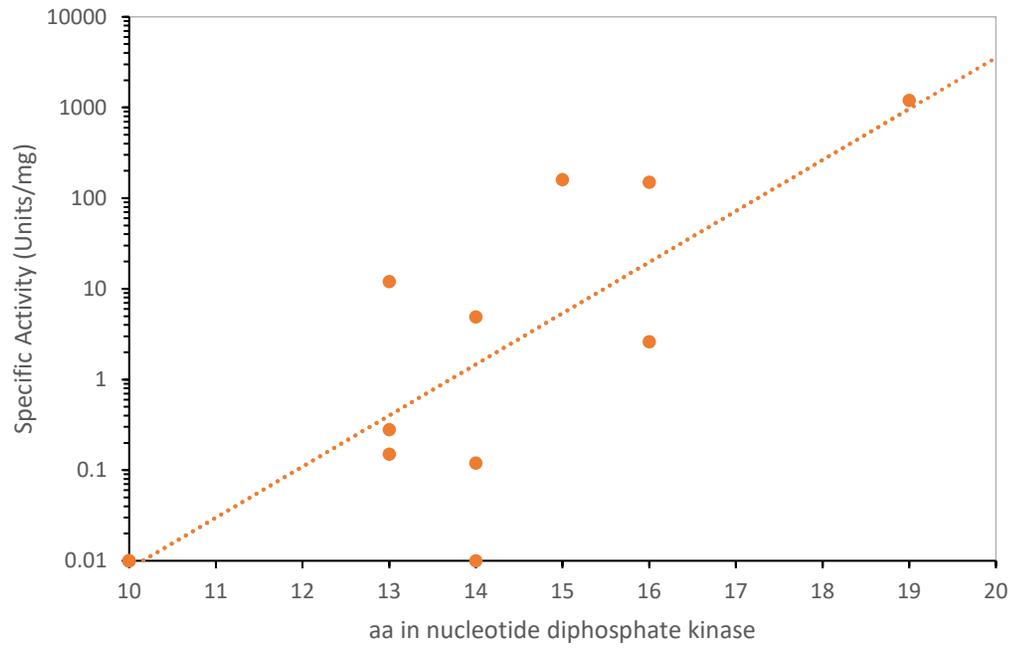

Figure 8